\documentclass[12pt]{article}
\usepackage{amsmath,amssymb,amsthm,bbm,latexsym}
\usepackage{graphicx,color}
\usepackage{xspace}
\usepackage[english]{babel}

\newcommand{\bee}{\begin{equation}}
\newcommand{\ee}{\end{equation}}
\newcommand{\bqa}{\begin{eqnarray}}
\newcommand{\eqa}{\end{eqnarray}}
\newcommand{\bea}{\begin{eqnarray}}
\newcommand{\eea}{\end{eqnarray}}

\def\R{{\mathbbm R}}

\def\Tr{{\rm Tr}}

\let\epsilon=\varepsilon

\newtheorem{lemma}{Lemma}

\newtheorem{theorem}{Theorem}

\newcommand{\cT}{{\cal T}}
\newcommand{\cG}{{\cal G}}

\begin{document}

\begin{titlepage}
\begin{flushright}
pi-qg-XXX\\
Lpt-Orsay-XXX
\end{flushright}

\vspace{20pt}

\begin{center}

{\Large\bf Spheres are rare}
\vspace{15pt}

{\large Vincent Rivasseau$^{a,\ddag} $}

\vspace{15pt}

$^{a}${\sl Laboratoire de Physique Th\'eorique, CNRS UMR 8627}\\
{\sl Universit\'e Paris-Sud, 91405 Orsay, France\\
and Perimeter Institute, Waterloo, Canada}
\vspace{5pt}

E-mail:  {\sl $^\ddag$rivass@th.u-psud.fr}

\vspace{10pt}

\begin{abstract}
We prove that triangulations of homology spheres in any dimension grow much slower than general
triangulations. Our bound states in particular that the number of triangulations 
of homology spheres in 3 dimensions grows at most like the power 1/3 of the 
number of general triangulations.
\end{abstract}
\end{center}

\noindent  Pacs numbers:  11.10.Gh, 04.60.-m
\\

\end{titlepage}

\section{Introduction}

The "Gromov question" \cite{Gromov} asks whether in dimensions higher than 2 the number of
triangulations of the sphere grows exponentially in the number of glued simplices,
as happens in dimension 2, for which explicit formulas are known \cite{Tutte,BIPZ,schaeffer}.
It has not been answered until now \cite{DJ,PZ,ColletEckmann}. It is usually formulated for triangulations that are 
\emph{homeomorphic} to a sphere. But we do not know counterexamples showing
that such an exponential bound could not hold also more generally for \emph{homology} spheres,
although we are conscious that homotopy constraints are much stronger than homology constraints.

Understanding general triangulations is important in the quantum gravity \cite{ADJ,Rov,GFZ} context.
Recently a theory of general (unsymmetrized) random tensors of rank $d$ was developped 
\cite{Gur1,Gur2,GurRyan1,univ,uncoloring}, with a new kind of $1/N$ expansion discovered \cite{Gur3,GurRiv,Gur4}.
This expansion is indexed by an integer, called the \emph{degree}. It is not a topological invariant but a sum of genera of 
\emph{jackets}, which are ribbon graphs embedded in the tensor graphs.
It also allowed to find an associated critical behavior \cite{riello} and to discover and study new classes
of renormalizable quantum field theories of the tensorial type \cite{BGR,BGLiv,cor1,VTD,cor2}.

In this note we perform a small step towards applying this new circle of ideas to the Gromov question. 
We prove a rather obvious result that we nevertheless could not find in the existing literature, 
namely that spherical triangulations are rare among all triangulations in any dimension. More precisely, we give in section 2  a 
necessary condition for a colored triangulation $\Gamma$ to have a trivial homology. 
It states that the rank of the incidence matrix of edges and faces for the dual graph $G$ of the triangulation $\Gamma$ must be equal
to the nullity of that graph (the number of edges not in a spanning tree).
From this condition we deduce that any such graph has always at least one jacket of relatively low genus.
 
In section 3 we prove that ribbon graphs with such relatively low genus are quite rare among general graphs. 
Combining the two results proves the statement of the title.

In particular in $d=3$ our bound states that triangulations of homology spheres made of $n$ tetrahedra grow at most as $ (n!)^{1/3}$,
while general triangulations made of $n$ tetrahedra grow as $n!$ (up to $K^n$ factors). 
Hence in dimension 3 spherical triangulations cannot grow faster than the cubic root of general triangulations.

An outlook of the connection with the tensor program for quantum gravity is provided in the last section.

\section{Spherical Triangulations}

To any ordinary triangulation is associated a unique colored triangulation, namely its barycentric subdivision.
If the initial triangulation is made of $n$ simplices of dimension $d$ (i.e. with $d+1$ summits), 
the barycentric subdivision is made of $a.n$
colored simplices, where $a$ only depends on the dimension $d$. For instance in $d=3$,
a tetrahedron is decomposed by barycentric subdivision (see Figure below) into 24 colored tetrahedra, hence $a=24$.
This map is  injective since the initial triangulation is made of the vertices of a single color, together
with lines that correspond to given bicolored lines of the colored triangulation.

\centerline{{\includegraphics[width=5cm,angle=0]{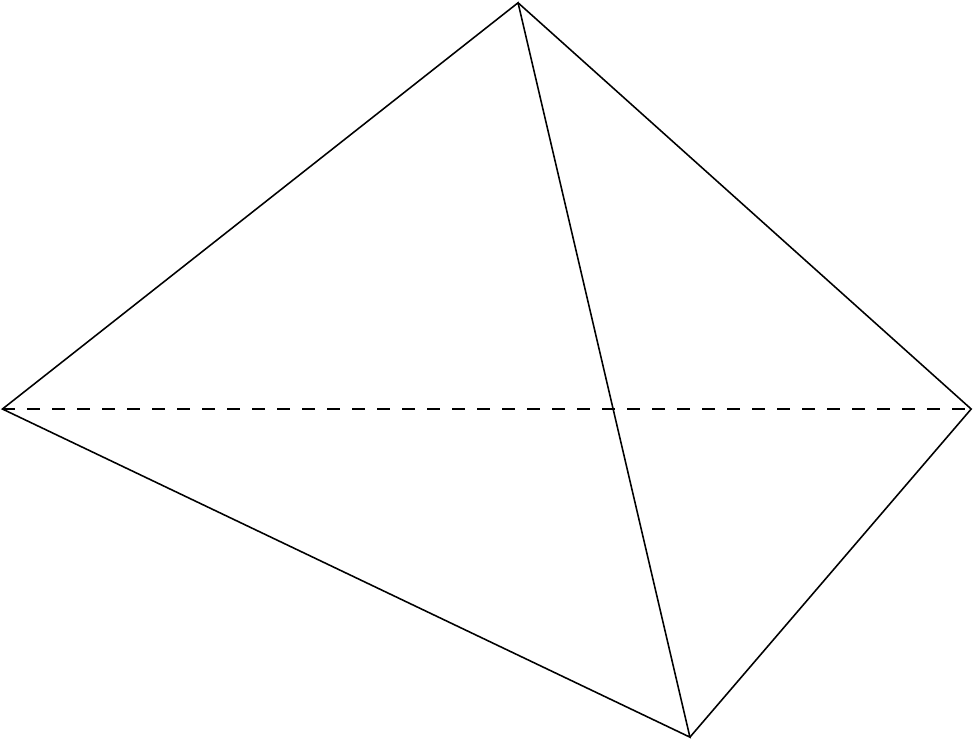}} \hskip1cm {\includegraphics[width=5cm,angle=0]{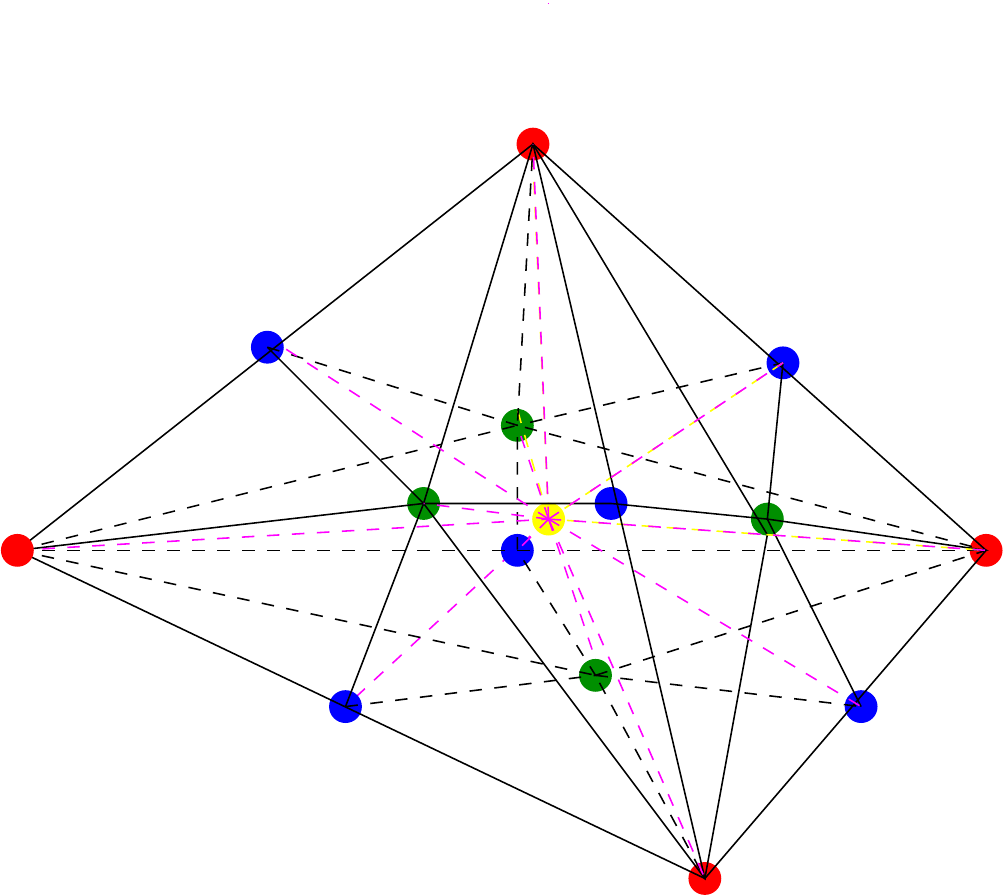}}}

Hence in order to study the Gromov question, one can restricts oneself to 
\emph{colored} triangulations. A bound in $K^n$ for spherical colored triangulations in $d=3$
would translate into a bound in $K^{24n}$ for spherical ordinary  triangulations and so on.

Colored triangulations \cite{FG,Lins} triangulate pseudo-manifolds \cite{Gur2}, and in contrast with the usual definition of spin foams
in loop quantum gravity, they have a well-defined $d$-dimensional homology. They are in one-to-one correspondence with 
dual \emph{edge-colored graphs}. Bipartite graphs correspond to triangulations of orientable pseudo-manifolds. 
Since they are associated to simple field theories \cite{Gur1} we expect they are the most natural objects for quantum gravity.


Therefore we consider from now on the category of (vacuum) connected bipartite edge-colored graphs with $d+1$ colors and 
uniform coordination $d+1$ (one edge of every color) at each vertex. 
The Gromov question can be rephrased as whether
the number of such graphs with $n$ vertices dual to spherical colored triangulations 
is bounded by $K^n$.

We call $V$, $E$ and $F$ the set of vertices, edges and faces of the graph $G$.
Faces are simply defined as the two-colored connected components of the graph, hence come in $d(d+1)/2$
different types, the number of pairs of colors. We also put $\vert V\vert =n $. $n$ is the order of the graph and is even since the graph is bipartite. Also
the graph being bipartite is naturally directed (i.e. there is a canonical orientation of each edge). We write $\cT$
for a generic spanning tree of $G$ and $G/\cT$ for the contracted graph with one vertex also called the \emph{rosette} associated to $G$ and $\cT$. 
The nullity of $G$ (number of loops, namely number of edges in any rosette) is $\vert L \vert = \vert E\vert -\vert V\vert  +1 = 1+ n(d-1)/2$.
 Jackets are ribbon graphs passing through all vertices of the graph.
There are $(d!)/2$ such jackets in dimension $d$ \cite{GurRiv}. Any jacket of genus $g$ provides a Hegaard decomposition
of the triangulated pseudo-manifold. In dimension 3 it gives a decomposition into two handle-bodies bounded by a 
common genus $g$ surface \cite{ryan}.

The edge-face incidence matrix $\epsilon_{ef}$ describes the incidence relation between edges and faces. Let us orient
each face arbitrarily: $\epsilon_{ef}$ is then $+1$, -1 or 0 depending on wether the face goes through the edge in the 
direct sense, opposite sense or does not go through $e$.\footnote{Faces running several times through an edge 
are excluded in colored graphs.}

Group field theory \cite{Boul,freidel,Oriti1} can be used to write connections on $G$ with a structure group $\cG$. 
To each edge of a group field theory graph $G$ is associated a generator $h_e \in \cG$ representing parallel transport along $e$.
The curvatures of the connection is the family of group elements $ \left[\overrightarrow{\prod_{e \in f}} {h_e}^{\epsilon_{ef}}\right] $,
for all faces $f \in F$, where the product is taken in the right order of the face.

The generators $h_e$ for the edges $e$ in any given spanning tree $\cT$ of $G$ 
are irrelevant for the computation of $\pi_1 (G)$, as they can be fixed to 1
through the usual  $G^{\vert V\vert}$ gauge invariance on connections.
Indeed setting $h_e =  1 \ \forall e \in \cT$ is equivalent to consider the 
retract $G/\cT$ of $G$ with a single vertex, hence the rosette associated to $G$ and $\cT$.\footnote{After this fixing 
the gauge transformations are reduced to a single global conjugation of all remaining $h_e$ by $G$.}

The fundamental group $\pi_1 (G)$ of $G$ admits then a presentation with one such generator $h_e$ per edge of
$G/\cT$ and the relations
\begin{equation}  \left[\overrightarrow{\prod_{e \in f}} {h_e}^{\epsilon_{ef}}\right]   = 1 \quad  \forall f\in F .  \label{fundgroup}
\end{equation}
Hence the space of flat connections (for which curvature is 1 for all faces) is the representation variety of $\pi_1 (G)$ into $G$
\cite{BS1}. 

In particular $G$ is simply connected if and only if the set of flat connections is just a point 
hence if the set of equations \eqref{fundgroup} has $h_e = 1 \  \forall e \in G/\cT $ as its unique solution.

The homology of $G$ is even simpler, as it corresponds to the case of a commutative group $\cG$. 
$G$ has trivial first-homology (i.e  $H_1 (G)=0$) if and only if
the set of commutative equations
\begin{equation}  \sum_{e \in L=G/\cT} \epsilon_{ef} h_e  = 0 \quad  \forall f\in F \label{firsthom}
\end{equation}
have $h_e = 0$ as unique solution.

By (commutative) gauge invariance the rank $r_G$ of the matrix $\epsilon_{ef}$ is equal to the rank of the reduced matrix $\epsilon^\cT_{ef}$
where the edges $e$ run over the reduced set of edges of $G/\cT$. We have certainly
\bee   r_G \le \inf \{ \vert  L \vert  , \vert  F\vert  \} .  \label{maxrank}
\ee

\begin{lemma}
The edge-colored graph $G$ has trivial first-homology if and only if
the rank of  $\epsilon^\cT_{ef}$ is maximal, i.e. equal to $\vert L \vert = 1 + (d-1)n/2 $. This writes:
\bee   r_G = 1 + (d-1)n/2  \label{rc}
\ee
\end{lemma}
\proof The rank $r_G$ cannot be larger than $\vert L \vert $ by \eqref{maxrank}. If it is strictly smaller it would mean that the linear map 
from $\R^L$ to $\R^F$ represented by the matrix $\epsilon_{ef}$ would have a non trivial kernel, hence the relations \eqref{firsthom}
defining the (first) homology space $H_1 (G)$ would have non-trivial solutions.
\qed

Remark that the rank condition \eqref{rc} implies that $\vert F \vert $ must be at least $\vert L \vert $ hence at least  $1 +(d-1)n/2 $.

In dimension $d$ we have $(d!)/2$ jackets $J$ and for each of them the relation
\bee  2 - 2 g(J) = n - \vert E\vert + F_J \implies g(J) =  1+ (d-1)n/4  - F_J/2
\ee 
 
Since each face belongs to exactly $(d-1)!$ jackets we have $\sum_J   F_J   =  (d-1)! \vert F\vert $ and the degree
of the graph is
\bee  \omega (G) = \sum_J g(J) = (d-1)! [ d/2+ d(d-1)n/8  -  \vert F\vert /2 ]
\ee 

It means that for a graph $G$ with $H_1 (G) = 0$, the degree obeys the bound
\bee 0 \le \omega (G)  \le   (d-1!) [ (d-1)/2+ (d -1)(d-2)n/8  ] . \label{degreebound}
\ee

In dimension $d$, duality exchanges the $k$-th and $(d-k)$-th Betti numbers.
\begin{lemma} 
In a graph $G$ with $H_1 (G) =0 $, hence dual to a triangulation $\Gamma$ such that $H_{d-1} (\Gamma) = 0$, there exists at least 
one jacket $J_0$ whose genus is bounded by
\bee   g(J_0)  \le \frac{d-1}{d}  [ 1 + \frac{(d-2)n}{4}  ] 
\ee
\end{lemma}
\proof We just divide the bound \eqref{degreebound} by the number $(d!)/2$  of the jackets.
\qed

Since spheres have trivial homologies hence zero Betti numbers between 1 and $d-1$, it follows that 
graphs dual to spherical triangulations all obey Lemma 2.
Of course triangulations of true (homotopy) spheres could be much rarer
but we won't investigate this question here.

\section{Low Genus Bounds}

In the previous section we proved that colored graphs 
dual to spherical triangulations must have at least
one jacket of relatively low genus. Let us now exploit this condition to bound the number of such graphs.
They are Feynman graphs and occur with their correct weights in the perturbative expansion
of a random tensor theory with action
\bee  e^{\lambda T_0 \cdots T_d  + \bar \lambda \bar T_0 \cdots \bar T_d   - \sum_{i=0}^d T_i \bar T_i} \label{tensor} .
\ee
where $T$ and $\bar T$ are tensors whose indices are contracted according to the pattern of the complete graph
on $d+1$ vertices. This is detailed at length in \cite{Gur1,GurRyan1}.

Consider now a particular jacket $J_0$. Suppressing all strands not in that jacket reduces
the tensorial  action to a matrix action of the type
\bee  e^{\lambda \Tr M_0 \cdots M_d  + \bar \lambda \Tr\bar M_0 \cdots \bar M_d   - \sum_{i=0}^d \Tr M_i M_i^\dagger}  . \label{matrix}
\ee

We know from Euler's formula that the corresponding  ribbon  Feynman graphs with $n$ 
vertices have genus $g$ bounded by $g \le g_{max} (n) = I(  \frac{2 + n(d-1)} {4})$, where $I$ means "integer part".
For $d=3$ this means that the genus of a bipartite 
ribbon graph of the $\phi^4$ type is bounded by 
$n/2$, where $n$ is the (even) order of the graph.

Let $T_{d,g,n}$ be the number of ribbon graphs  of order $n$ and genus $g$ corresponding to the action \eqref{matrix}.
Our main bound is
\begin{lemma}
There exists a constant $K_d$ such that
\bee \vert T_{d,g,n} \vert   \le    K_d^n  n^{2g}  \label{uppercomb}
\ee 
\end{lemma}
\proof
Because of the bipartite character of action \eqref{matrix}, $n=2p$ is even.
We want to count the number of Wick contractions matching $4p$ fields and $4p$ anti-fields on $p$ vertices and $p$ 
anti-vertices giving rise to a ribbon graph of genus $g$. 
The edges of such a graph can always be decomposed into a spanning tree $\cT$ of $n-1$ edges, a dual 
tree $\tilde \cT$ in the dual graph made of $\vert F\vert -1  =  n+1-2g$ edges, and a set of $2g$ "crossing edges" $CE$.\footnote{There are 
usually many different such decompositions but we just choose arbitrarily one of them for each graph.} Paying an overall factor
$3^{4n}$ we can preselect as $(A, \bar A)$, $(B, \bar B)$ and $(C, \bar C)$
the fields and anti-fields which Wick-contract respectively into $\cT$, $\tilde \cT$ and $CE$. Building the Wick contractions between 
the $(n-1)$ fields of $A$ and the $n-1$ anti-fields of $\bar A$ to form the labeled tree $\cT$ certainly costs at most $(n-1)!$, the total
number of such contractions. Contracting the tree $\cT$ to a single vertex we obtain a cyclic ordering
of the remaining $2(n+1)$ fields and anti-fields of $B\cup \bar B \cup C \cup \bar C$. Let us delete 
for the moment on the cycle the $2g$ fields of $C$ and the $2g$ anti-fields of $\bar C$. 
Building the dual tree $\tilde \cT$ out of contractions of the $n+1-2g $ fields of $B$ and $n+1-2g$ antifields of $\bar B$
must create a new face per edge, hence the number of corresponding Wick contractions is bounded 
by the number of non-crossing matchings between $B$ and $\bar B$ on
the cycle. We know that the total number of such non-crossing matchings between $2p$ 
objects is the Catalan number $C_p \le 4^p$, hence we obtain a bound $4^{n+1}$ for the Wick contractions of the fields of $B$ 
with the anti-fields  of $\bar B$. Finally the number of contractions joining the $2g$ fields of $C$ to the $2g$ anti-fields of $\bar C$ to create
the edges of $CE$ can be bounded by joining them in any possible way, hence by $(2g)!$. 
Using the standard vertex symmetry factor $[p!]^{-2}$ of Feynman graphs coming from expanding the exponential action in \eqref{matrix}
(and since $g \le g_{max} (n) =  I(  \frac{2 + n(d-1)} {4})$), we 
easily conclude that building $\cT$,  $\tilde \cT$ and $CE$ costs at most $K^n  n^{2g} $ Wick contractions, and we get \eqref{uppercomb}
with $K_d = 3^4 K$.
\qed

Notice that we did not try at all to optimize $K_d$ (in particular in the proof of the Lemma above we did not try to use the colors which give further
constraints, as they would not improve on the factor $n^{2g}$. 
Remark also that the upper bound \eqref{uppercomb} of Lemma 3 does not contradict  
the well-established apparently larger asymptotic behavior \cite{CMS}
\bee  T_{g,n}  \simeq_{n \to \infty}   c_g  \cdot n^{5/2 (g-1)}  12^n
\ee
for \emph{fixed} $g$ and large $n$ which is e.g. used to define double scaling in matrix models. Indeed this 
asymptotic behavior cannot be maintained when $g$ grows with $n$, as we know that $g \le n/2$.

Let us neglect fixed powers of $n$ since they can be absorbed into new $K^n$ factors.
In dimension $d$ general connected graphs at order $n$ grow
as $K^n  n^{n(d-1)/2}$, as expected for a $\phi^{d+1}$ interaction. The number of
graphs satisfying Lemma 2 on the other hand is bounded by $(d!)/2$ (to choose the jacket $J_0$) times
the number of ribbon graphs with genus $g \le \frac{d-1}{d}  [ 1 + \frac{(d-2)n}{4}  ] $ in that $J_0$ jacket. By Lemma 3 it is therefore bounded by
$(K')^n \cdot n^{2g (J_0)}$, hence by $(K")^n  \cdot n^{\frac{(d-1)(d-2)}{2d}n}$.

Putting together these results we obtain
\begin{theorem}
There exist constants $K$ and $K'$ such that the number $ST_n$ of spherical triangulations with $n$ simplices is bounded by
\bee ST_N  \le K^n n^{\frac{(d-1)(d-2)}{2d}n}
\ee
and such that the number of general triangulations $T_n$ obeys
\bee T_N  \ge (K')^n n^{\frac{d-1}{2}}
\ee
Since in any dimension $\frac{(d-1)(d-2)}{2d}  < \frac{d-1}{2}$, we have 
\bee  \lim_{n \to \infty}  ST_n/T_n  =0
\ee
which means that triangulations of spheres are always rare among general triangulations.
\end{theorem}

In dimension 3 we get that triangulations of homology spheres grow at most as $(n!)^{1/3}$ whether
general triangulations grow at least as $n!$. Hence spherical triangulations cannot grow faster than a cubic root 
of general triangulations.

\section{Outlook}  We would like a microscopic theory of quantum gravity to sum over all spaces irrespectively of their topology and
to generate  a macroscopic space-time such as the one we observe (large and of trivial topology).
Since spheres are rare, this cannot be done without some non-trivial ponderation
factor to favor them. Random tensor models have precisely such a factor; they ponder triangulations not by 1
but by $\lambda^{\vert V\vert} N^{\vert F\vert}$ where $\lambda$ plays the role of the cosmological constant, 
$N$ is the size of the tensor and $N^{\vert F\vert}$
is a discretization of the Einstein-Hilbert action. Indeed for flat equilateral triangulations, curvature is concentrated on the $d-2$
dimensional simplices, hence is associated to the faces of the dual graphs.  

The $1/N$ expansion of random tensor models has melons (i.e. very particular "stacked" triangulations of the spheres)
as their leading graphs. It can therefore lead from a perturbative phase around "no space at all" to the condensation of a primitive
kind of space-time, namely the continuous random tree (CRT) \cite{aldous}, called branched polymer by physicists. Indeed melons are 
branched polymers \cite{gurauryan}. This melonic CRT phase  (of Hausdorff dimension 2) however \emph{cannot be the end of the 
tensor story}. Indeed sub-melonic triangulations include e.g. at least all graphs planar  in a fixed jacket. Such graphs
are exactly well-labeled trees \cite{schaeffer}. The labels 
create shortcuts on the CRT, leading to the very different 2D Brownian sphere phase 
(of Hausdorff dimension 4 \cite{CS}).\footnote{They also realize a nice concrete toy model for the holographic 
principle, since all information about the faces of the triangulation is captured by the labels on 
the tree which has a single face as its boundary.} We hope that investigation of all submelonic
contributions in higher rank random tensors will uncover similar but more complicated shortcuts on the melonic CRT. 
Ideally it could then lead through a sequence of phase transitions to an effective space-time similar to 
the one we observe, in which e.g. topological, spectral and Hausdorff dimensions all 
appear equal to 4.

However even if this is the case, many mysteries would remain. Let us briefly discuss one of them.
The two dimensional phase transition from planar graphs to Brownian spheres or 
the higher-dimensional phase transition from melons to the continuous random tree occurs for the \emph{unstable} sign of the coupling constant $\lambda$, where
all graphs add up with the same sign. But we want to (Borel)-sum \emph{all} triangulations (not only planar or melonic ones).
Random tensors models can do that using the loop vertex expansion \cite{LVE,univ,MNRS}, 
but only for the \emph{other sign} of the coupling constant, in which amplitudes alternate with their order.
It is also for this stable sign of the coupling constant that renormalizable tensor group field theories \cite{BGR,BGLiv, cor1,VTD,cor2}
have been proved asymptotically free \cite{bengeloun}, meaning the coupling constant $\lambda$ in such theories 
grows naturally and should unavoidably reach some critical value and generate a phase transition. 

It is tempting but difficult to integrate all these insights into a single coherent picture, ideally that of 
a renormalizable tensor group field theory whose renormalization group trajectory would lead
through a sequence of phase transitions from no space at all to the 4D space we know of, equipped with general relativity
as an effective theory. Indeed a major difficulty -underlined e.g. by Ambjorn \cite{Amb}- is this incoherence of sign: Borel summability and asymptotic
freedom require one sign of the coupling, when the CRT and Brownian sphere phase transitions occur for the other sign.\footnote{
This difficulty also is the main reason for which double scaling in matrix models
is unstable.} 
To connect both phenomena seems to require some kind 
of analytic continuation; let us simply remark that coupling constant flows which grow in the infrared 
are unstable under addition of an infinitesimal imaginary part, 
which can send them to the other side of the real axis. One could also speculate that it is perhaps at this point that it may be necessary to 
supplement the purely Euclidean approach with some form of continuation to a Lorentzian metric. 

Answering the Gromov question is not a prerequisite for the tensor track program \cite{Track1,Track2}, which proposes to 
ultimately (Borel)-sum over all triangulations anyway. However it is unclear whether submelonic corrections and
the precise geometrogenetic phase transitions they could generate can be investigated in detail if we 
remain unable to answer the relatively simple and natural Gromov question. A positive
answer would reinforce the analogy between planar graphs and random matrices on one side and spherical triangulations and random tensors
on the other. It would lend some weight to the hope that e.g. the equally weighted measure on spherical triangulations in dimensions
3 and 4 could converge (in the Gromov-Hausdorff sense) to a new kind of compact random space generalizing
the Brownian two-dimensional sphere \cite{LeGall,LeGallMiermont,Miermont} to higher dimensions. Also it would open the possibility 
that double or multiple scalings beyond the melonic graphs could be found \emph{within}
the spherical triangulations; in that case we would expect the result to be stable, in contrast to what happens
in matrix models.


\medskip
\noindent{\bf Acknowledgments} 
I thank R. Gurau for useful discussions. Research at Perimeter Institute is supported by the Government of Canada through 
Industry Canada and by the Province of
Ontario through the Ministry of Research and Innovation


\begin{thebibliography}{99}

\bibitem{Gromov} M. Gromov. 
Spaces and questions. Geom. Funct. Anal. (2000), 118Ð161GAFA 2000 (Tel Aviv, 1999).

\bibitem{Tutte}
W. T. Tutte. A census of planar triangulations. Canad. J. Math. 14 (1962), 21Ð38.
    
\bibitem{BIPZ}  
E. Br\'ezin, C. Itzykson, G. Parisi, and J. B. Zuber,  Planar diagrams,
Comm. Math. Phys. {\bf 59}, (1978), 35-51. 

  \bibitem{schaeffer}  
G. Schaeffer, Conjugaison d'arbres et cartes combinatoires al\'eatoires, PhD thesis 1998.
\bibitem{DJ}
B. Durhuus and T. J\'onsson, Remarks on the entropy of 3-manifolds. Nuclear Phys. B 445 (1995),
182Ð192.

\bibitem{PZ} 
J. Pfeiße and G. M. Ziegler, Many triangulated 3-spheres. Math. Ann. 330 (2004), 829Ð837.

\bibitem{ColletEckmann}
P Collet, J.-P. Eckmann, M. Younan,
Trees of nuclei and bounds on the number of triangulations of the 3-ball,
arXiv:1204.6161
 
   
\bibitem{ADJ}
J. Ambj¿rn, B. Durhuus, and T. Jonsson. Quantum geometry. Cambridge Monographs on Mathematical
Physics (Cambridge: Cambridge University Press, 1997).


\bibitem{Rov}
C. Rovelli, Quantum Gravity, Cambridge University Press (2004).


\bibitem{GFZ} P. Di Francesco. P. Ginsparg and J. Zinn-Justin. 2D Gravity and Random Matrices. Physics Reports {\bf 254},
1-131 (1995).

\bibitem{Gur1}
 R.~Gurau,
  ``Colored Group Field Theory,''
  [arXiv:0907.2582 [hep-th]].

\bibitem{Gur2}
  R.~Gurau,
  ``Lost in Translation: Topological Singularities in Group Field Theory,''
  Class.\ Quant.\ Grav.\  {\bf 27}, 235023 (2010)
  [arXiv:1006.0714 [hep-th]].
  
\bibitem{GurRyan1} 
  R.~Gurau and J.~P.~Ryan,
  SIGMA {\bf 8}, 020 (2012)
  [arXiv:1109.4812 [hep-th]].
   
\bibitem{univ} 
  R.~Gurau,
  ``Universality for Random Tensors,''
  arXiv:1111.0519 [math.PR].
  
\bibitem{uncoloring} 
  V.~Bonzom, R.~Gurau and V.~Rivasseau,
  ``Random tensor models in the large N limit: Uncoloring the colored tensor models,''
  Phys.\ Rev.\ D {\bf 85}, 084037 (2012)
  [arXiv:1202.3637 [hep-th]].

\bibitem{Gur3}
R.~Gurau,
``The 1/N expansion of colored tensor models,''
arXiv:1011.2726 [gr-qc].

\bibitem{GurRiv}
   R.~Gurau and V.~Rivasseau,
  ``The 1/N expansion of colored tensor models in arbitrary dimension,''
  arXiv:1101.4182 [gr-qc].

\bibitem{Gur4}
  R.~Gurau,
  ``The complete 1/N expansion of colored tensor models in arbitrary
  dimension,''
  arXiv:1102.5759 [gr-qc].
  
\bibitem{riello} 
  V.~Bonzom, R.~Gurau, A.~Riello and V.~Rivasseau,
  ``Critical behavior of colored tensor models in the large N limit,''
  Nucl.\ Phys.\ B {\bf 853}, 174 (2011)
  [arXiv:1105.3122 [hep-th]].

\bibitem{BGR}
  J.~Ben Geloun and V.~Rivasseau,
  ``A Renormalizable 4-Dimensional Tensor Field Theory,''
  Commun.\ Math.\ Phys.\  {\bf 318}, 69 (2013)
  [arXiv:1111.4997 [hep-th]].

\bibitem{BGLiv}
  J.~B.~Geloun and E.~R.~Livine,
  ``Some classes of renormalizable tensor models,''
  arXiv:1207.0416 [hep-th].

\bibitem{cor1} 
  S.~Carrozza, D.~Oriti and V.~Rivasseau,
  ``Renormalization of Tensorial Group Field Theories: Abelian U(1) Models in Four Dimensions,''
  arXiv:1207.6734 [hep-th].
  
  \bibitem{VTD} 
  D.~O.~Samary and F.~Vignes-Tourneret,
  ``Just Renormalizable TGFT's on $U(1)^{d}$ with Gauge Invariance,''
  arXiv:1211.2618 [hep-th].

\bibitem{cor2}   S.~Carrozza, D.~Oriti and V.~Rivasseau, 
Renormalization of an SU(2) Tensorial Group Field Theory
in Three Dimensions, in preparation

\bibitem{FG}
  M. ~Ferri and C.~Gagliardi
  Pacific\ Journal\ of\ Mathematics\ Vol. 100, No. 1, 1982

\bibitem{Lins} S.~Lins, 
  ISBN: 9810219075/ ISBN-13: 9789810219079 
 
  
\bibitem{ryan}
J. Ryan,  Tensor models and embedded Riemann surfaces, arXiv:1104.5471   


\bibitem{Boul}
D. V Boulatov, Mod.Phys.Lett. A7:1629-1646 (1992), [arXiv:hep-th/9202074]

\bibitem{freidel}
L. Freidel, Int.J.Phys. 44, 1769-1783, (2005) [arXiv: hep-th/0505016]

\bibitem{Oriti1} 
  D.~Oriti,
 ``The microscopic dynamics of quantum space as a group field theory,''
  arXiv:1110.5606 [hep-th].
   
 \bibitem{BS1}
  V.~Bonzom and M.~Smerlak,
  ``Bubble divergences from cellular cohomology,''
  Lett.\ Math.\ Phys.\  {\bf 93}, 295 (2010)
  [arXiv:1004.5196 [gr-qc]].

\bibitem{CMS}
 G. Chapuy, M. Marcus, G. Schaeffer, SIAM Journal on Discrete Mathematics, 23(3):1587-1611 (2009)
 A bijection for rooted maps on orientable surfaces, arXiv:0712.3649
 
 
\bibitem{aldous} D. Aldous, The Continuum Random Tree I, II and III; The Annals of Probability, 1991, Vol 19, 1-28; in 
Stochastic Analysis, London Math Society Lecture Notes, Cambridge University Press 1991n eds Barlow and Bingham;
The Annals of Probability, 1993, Vol 21, 248-289.




 \bibitem{gurauryan}
  R.~Gurau and J.~P.~Ryan,
  ``Melons are branched polymers,''
  arXiv:1302.4386 [math-ph].
  
  \bibitem{CS}
P. Chassaing and G. Schaeffer,
Random Planar Lattices and Integrated
SuperBrownian Excursion, Probability Theory and Related Fields, 128(2):161--212.
   

 \bibitem{LVE} 
  V.~Rivasseau,
 ``Constructive Matrix Theory,''
  JHEP {\bf 0709}, 008 (2007)
  [arXiv:0706.1224 [hep-th]].
  
  
\bibitem{MNRS} 
  J.~Magnen, K.~Noui, V.~Rivasseau and M.~Smerlak,
 ``Scaling behaviour of three-dimensional group field theory,''
  Class.\ Quant.\ Grav.\  {\bf 26}, 185012 (2009)
  [arXiv:0906.5477 [hep-th]].
   
   
   
\bibitem{bengeloun}
  J.~Ben Geloun,
 ``Two and four-loop $\beta$-functions of rank 4 renormalizable tensor field theories,''
  Class.\ Quant.\ Grav.\  {\bf 29}, 235011 (2012)
  [arXiv:1205.5513 [hep-th]].
  
 
 \bibitem{Amb} 
 J.~Ambjorn, private communication.
 

   \bibitem{Track1} 
  V.~Rivasseau,
``Quantum Gravity and Renormalization: The Tensor Track,''
  AIP Conf.\ Proc.\  {\bf 1444}, 18 (2011)
  [arXiv:1112.5104 [hep-th]].
  
\bibitem{Track2} 
  V.~Rivasseau,
  ``The Tensor Track: an Update,''
  arXiv:1209.5284 [hep-th].
  
  \bibitem{LeGall}   J.F. Le Gall, 
Uniqueness and universality of the Brownian map,
 arXiv:1105.4842

\bibitem{LeGallMiermont} J.-F. Le Gall, G. Miermont,
Scaling limits of random trees and planar maps,
arXiv:1101.4856

\bibitem{Miermont}  G. Miermont,
The Brownian map is the scaling limit of uniform random plane quadrangulations,
arXiv:1104.1606




\end{thebibliography}
\end{document}